\documentclass{article}
\usepackage{spconf,amsmath,graphicx,makecell,amsfonts}

\title{dimensionality reduction via diffusion map improved \\ with supervised linear projection}
\name{Bowen Jiang$^{\dagger}$, Maohao Shen$^{\dagger}$ \thanks{$^{\dagger}$These two authors contribute equally to the work.}}
\address{Department of Electrical and Computer Engineering \\
University of Illinois Urbana-Champaign, IL, USA \\
\texttt{\small\{bowenj2,maohaos2\}@illinois.edu}
}
\begin{document}

$
\begin{array}{l}\huge{\textbf {IEEE Copyright Notice }}\\ \\ \\ \copyright \ \text {2020 IEEE. } \text { Personal use of this material is permitted. Permission from IEEE must be obtained for all other } \\ \text {uses, in any current or future media, including reprinting/republishing this material for advertising or promotional } \\ \text {purposes, creating new collective works for resale or redistribution to servers or lists, or reuse of any copyrighted } \\ \text {component of this work in other works. } \\ \\ \\ \text {Accepted to be published in: The 27th IEEE International Conference on Image Processing (ICIP 2020) } \\ \text {Abu Dhabi, United Arab Emirates (UAE), October 25-28, 2020. }\end{array}
$

\topmargin=0mm
\ninept
\maketitle
\begin{abstract}
When performing classification tasks, raw high dimensional features often contain redundant information, and lead to increased computational complexity and overfitting. In this paper, we assume the data samples lie on a single underlying smooth manifold, and define intra-class and inter-class similarities using pairwise local kernel distances. We aim to find a linear projection to maximize the intra-class similarities and minimize the inter-class similarities simultaneously, so that the projected low dimensional data has optimized pairwise distances based on the label information, which is more suitable for a Diffusion Map to do further dimensionality reduction. Numerical experiments on several benchmark datasets show that our proposed approaches are able to extract low dimensional discriminate features that could help us achieve higher classification accuracy.
\end{abstract}
\begin{keywords}
Dimensionality reduction, diffusion map, image and video classification, image processing, supervised learning
\end{keywords}
\section{Introduction}
\label{sec:intro}

Recent decades have witnessed great interest in the dimensionality reduction algorithms, which are applied on high dimensional data before image or video classification. To perform classification, useful discriminative features, rather than raw image pixels, should be extracted before dimensionality reductions. Unfortunately, however, these feature vectors tend to be overwhelmingly high dimensional with too much redundant information. This not only leads to increased computational complexity and the overfitting problem in the learning process, but also creates the famous ''curse of dimensionality'' problem \cite{Bellman:1957,trunk1979problem} where the amount of data required to perform a statistically stable analysis grows exponentially. That is why dimensionality reduction before classification is necessary.

We define the satisfactory dimensionality reduction as one that preserves or improves classification accuracy. Ideally, the data after dimensionality reduction should behave so that the samples belonging to the same class have similar attributes - more similarities - and the samples belonging to different classes have different attributes. Because only local data samples are considered, the similarity can be defined being proportional to the pairwise kernel $l_2$ distance of data samples as
\begin{equation}
\text{similarity between }\hat{\boldsymbol{x}_i}\text{ and }\hat{\boldsymbol{x}_j} \propto  \exp - \frac{\parallel \hat{\boldsymbol{x}_i} - \hat{\boldsymbol{x}_j} \parallel^2 _2}{\sigma^2}
\end{equation}
where $\hat{\boldsymbol{x}_i}, \hat{\boldsymbol{x}_j}$ are two arbitrary data samples after dimensionality reduction, $\sigma$ is the parameter related to the dataset variance, and the exponential term $\exp \left(-{\parallel \hat{\boldsymbol{x}_i} - \hat{\boldsymbol{x}_j} \parallel^2 _2}/{\sigma^2}\right)$ is defined as the kernel distance, which is inversely related to the Euclidean distance between $\hat{\boldsymbol{x}_i}$ and $\hat{\boldsymbol{x}_j}$. 

A variety of low dimensional feature extraction methods available for dimensionality reduction exists in the literature. For example, the Principal Component Analysis (PCA) \cite{jolliffe2002principal,abdi2010principal} linearly projects a high dimensional feature vector onto a set of orthonormal basis so that the sample variance is maximized along each dimension. However, PCA is an unsupervised method, and does not take account of the intra-class similarities and inter-class similarities. Linear Discriminant Analysis (LDA) \cite{fisher1936use,mclachlan2004discriminant} is more suited for extracting discriminative features. It tries to minimize the intra-class covariance and maximize the inter-class covariance at the same time while searching for the projection directions. When the data samples are normally distributed, LDA is guaranteed to produce the Bayes optimal solution \cite{mclachlan2004discriminant}. Since the intra-class covariance is computed using all the samples in the same class, it might not be the best measure to reflect intra-class similarities when the data follow some distributions other than the Gaussian case.

Here we consider a special type of data that contains underlying non-linear structures in its manifolds in the image space. Human action sequencing with smooth motion changes is a typical example \cite{gorelick2007actions}. In such cases, linear dimensionality reduction alone fails to preserve the local geometric information. In contrast, the Diffusion Map \cite{coifman2006diffusion}, as an unsupervised non-linear algorithm, has a well-known ability to discover the non-linear hidden manifold from this kind of high dimensional data, and to preserve the intrinsic low dimensional geometric shapes. This is not good enough, however, because a Diffusion Map uses no label information, and we believe that the label information given by the training dataset could be helpful. Therefore, for better classification purposes based on training data label information, we want to use a supervised linear projection to re-adjust data distributions before data is sent to the Diffusion Map, so that the intra-class similarities are maximized, while the inter-class similarities are minimized. Meanwhile, we also want dimensionality reduction in this supervised projection, so that the Diffusion Map can better discover the manifold with less influence from too much redundant information. 

Eventually, the classification accuracy performed by classifiers on the discovered manifold is expected to be higher than that without the proposed supervised linear projection. Various classical classifiers exist, including linear support vector machines (SVM) \cite{cortes1995support} and k-nearest neighbours (KNN).

The paper proceeds as follows: Introduction of objective function in Section 2. Details of supervised dimensionality reduction algorithms in Section 3. Discussion in Section 4. Experiment results in Section 5. Conclusion and future works in Section 6.

\section{OBJECTIVE FUNCTION}
\label{sec:objfunc}

We would like to apply a supervised linear projection on labeled training data $\boldsymbol{x} \in \mathbb{R}^D$ as inputs, trying to maximize the similarities, defined in (1), of data within the same label, while minimizing the similarities of data from different labels. Top eigenvectors and eigenvalues are selected from the optimization solution (9) mentioned in Section 3 for dimensionality reduction, and the low dimensional data is labeled as $\hat{\boldsymbol{x}} \in \mathbb{R}^d$ with $d < D$.

Let $\boldsymbol{x}$ be the data in the high dimensional space, which represents feature vectors extracted from original images using methods such as SIFT \cite{lowe1999object,vedaldi2010vlfeat}, $\boldsymbol{z}_{ij} = \boldsymbol{x}_i - \boldsymbol{x}_j$ be the distance between high dimensional data $\boldsymbol{x}_i$ and $\boldsymbol{x}_j$, $\hat{\boldsymbol{x}}$ be the low dimensional data after the linear projection $\boldsymbol{P}$, and $M_{ij} = \ \parallel \hat{\boldsymbol{x}_i} - \hat{\boldsymbol{x}_j} \parallel^2 _2 \ = \boldsymbol{z}_{ij}^{\top}\boldsymbol{PP}^{\top}\boldsymbol{z}_{ij} $ be distances between low dimensional data $\hat{\boldsymbol{x}}_i$ and $\hat{\boldsymbol{x}}_j$.

Using the Lagrange multipliers $\lambda$, the objective function is
\begin{equation}
\begin{aligned}
&\max_{\boldsymbol{P}}\ \ J(\boldsymbol{P}) - \lambda \left(\boldsymbol{P}^{\top}\boldsymbol{P}-\boldsymbol{I}\right)\\
&\text{where}\ \ J(\boldsymbol{P}) = \left( 1-\rho \right) \sum_{k=1}^{K} \lambda_k \sum_{i,j\in k} \exp^{-\frac{M_{ij}}{\sigma^2}} - \rho\sum_{i,j\notin k} \exp^{-\frac{M_{ij}}{\sigma^2}}
\end{aligned}   
\end{equation}
where the exponential term $\sum_{k=1}^{K} \lambda_k \sum_{i,j\in k} \exp\left(-M_{ij}/\sigma^2\right)$ represents the sum of kernel distances of data within each class, and over all classes. Different $\lambda_k$ allow different classes to have different weights. The other exponential term $\sum_{i,j\notin k} \exp\left(-M_{ij}/\sigma^2\right)$ is the sum of kernel distance of data from different classes. $\left(1-\rho \right)$ and $\rho$ balance the weights 
between intra-class and inter-class similarities.

Because the amount of data from the same class is usually much less than that from different classes, it is empirically recommended to use relatively large $\lambda_k$ and small $\rho$, so that the values of two summations of kernel distances from intra-class and inter-class samples in (2) can be less skewed and more balanced. 

\section{DIMENSIONALITY REDUCTION}
\label{sec:dimreduct}
\subsection{First Order Taylor Approximation}
\label{ssec:approx}

Because the objective function involves exponential terms, optimization by taking derivatives directly of (2) doesn't work. As the result, in order to maximize (2) with respect to projection matrix $\boldsymbol{P}$, we perform optimization similar to Newton's Method to embed $\boldsymbol{P}$ into the original objective function, so that a closed form solution (9) can be obtained.

Before Taylor approximation, to eliminate the exponential terms for calculation convenience, we rewrite the terms in (2) as
\begin{equation}
\begin{aligned}
&\begin{cases}
\sum_{k=1}^K \sum_{i,j \in k} \exp^{- \frac{M_{ij}}{\sigma^2}} = - \sum_{k=1}^K n_k \ln\left(m_{c_k}\right) \\
\sum_{i,j \notin k} \exp^{- \frac{M_{ij}}{\sigma^2}} = -n_o \ln\left(m_{o}\right)
\end{cases} \\
&\text{where}
\begin{cases}
\ m_{c_k} = \exp \left( -\frac{1}{n_k} \sum_{i,j \in k} \exp^{- \frac{M_{ij}}{\sigma^2} }\right) \\
\ m_{o} = \exp \left( -\frac{1}{n_o} \sum_{i,j \notin k} \exp^{- \frac{M_{ij}}{\sigma^2} }\right)
\end{cases}
\end{aligned}
\end{equation}
each $n_k$ is the total number of data pairs within the same class, and $n_o$ is the total number of data pairs among different classes.

Intuitively, regardless of additional logarithms and exponentials outside, $m_{c_k}$ and $m_o$ in (3) are positively correlated with the numerical averages of low dimensional Euclidean distance $M_{ij}$. On the other hand, the logarithms have nice derivative forms to simplify the calculations later in (4).

Therefore, with $t$ representing the current iteration, and $t+1$ the next iteration, the objective function (2) after the first order Taylor approximation becomes
\begin{equation}
\begin{aligned}
& J^{(t+1)}(\boldsymbol{P}) = \\
&  \left( 1-\rho \right) \sum_{k=1}^K \lambda_k \sum_{i,j \in k} \left( -\ln(m_{c_k}^{(t)}) - \frac{1}{m_{c_k}^{(t)}}  \left(\boldsymbol{z}_{ij}^{\top}\boldsymbol{PP}^{\top}\boldsymbol{z}_{ij}-m_{c_k}^{(t)}\right) \right) \\
& -\rho\sum_{i,j \notin k} \left(-\ln\left(m_o^{(t)}\right) - \frac{1}{m_o^{(t)}} \left(\boldsymbol{z}_{ij}^{\top}\boldsymbol{PP}^{\top}\boldsymbol{z}_{ij}-m_o^{(t)}\right) \right)
\end{aligned}
\end{equation}
\vspace{-0.5cm}

\subsection{Maximization of the Objective Function}
\label{ssec:optimize}
To find the objective function updating rule of each iteration, we optimize the objective function (4) by taking first order derivative,
\begin{equation}
\begin{aligned}
\frac{\partial J^{(t+1)}(\boldsymbol{P}) -\lambda\left(\boldsymbol{P}^{\top}\boldsymbol{P}-\boldsymbol{I}\right)}{\partial \boldsymbol{P}} = 0
\end{aligned}
\end{equation}
\begin{equation}
\begin{aligned}
&\Rightarrow \left(\sum_{i,j} \alpha_{ij}^{(t)} z_{ij}z_{ij}^{\top} \right) \boldsymbol{P} = \lambda \boldsymbol{P} 
\end{aligned}
\end{equation}
\begin{equation}
\text{where} \ \alpha_{ij}^{(t)} = \begin{cases}
-\frac{1}{m_{c_k}^{(t)}} \left(1-\rho\right)\ \lambda_k , & \text{if } i,j\in k \\
\frac{1}{m_{o}^{(t)}} \ \rho , & \text{if } i,j \notin k \ \ \ \ \ \ \ \ \ \ \ \ \ \ \ \ \ \ \ \ \ \ \ \ \ \ \ \ \ \ \ \ \ 
\end{cases}
\end{equation}

Given current $m_{c_k}^{(t)}$ and $m_o^{(t)}$ , as well as constant factors $\left(1-\rho\right)$ and $\lambda_k$ initialized at the beginning, we can obtain $\alpha_{ij}$ of the current iteration, which will be used for calculating the objective function of the next iteration.

Because the equation (6) can be further simplified as
\begin{equation}
\begin{aligned}
\sum_{i\ne j} \alpha_{ij}^{(t)} z_{ij}z_{ij}^{\top} &= 2 \sum_{i,j} x_i \alpha_{ij}^{(t)} x_i^{\top} - 2 \sum_{i,j} x_i \alpha_{ij}^{(t)} x_j^{\top}  \\
& = \ \boldsymbol{XEX}^{\top} \ - \ \boldsymbol{XDX}^{\top} \ = \ \boldsymbol{XLX}^{\top}
\end{aligned}
\end{equation}
where $E_{ii} = 2 \sum_{j}\alpha_{ij}^{(t)}$ is diagonal, $D_{ij} = 2\alpha_{ij}^{(t)}$, $\boldsymbol{L} = \boldsymbol{E} - \boldsymbol{D}$, and $\boldsymbol{X}$ is the matrix with each $\boldsymbol{x}$ as its columns, we get
\begin{equation}
\begin{cases}
\boldsymbol{P} = \text{top eigenvectors of} \left\{\boldsymbol{XLX}^{\top}\right\} \\
\lambda = \boldsymbol{P}^{\top}\boldsymbol{XLX}^{\top}\boldsymbol{P}
\end{cases}
\end{equation}

Assume we want to reduce the dimensionality to $d < D$, we would like to select the largest $d$ positive eigenvalues of $\boldsymbol{XLX}^{\top}$, and corresponding $d$ eigenvectors concurrently. Then each row of the projection matrix $\boldsymbol{P}$ is one of the selected eigenvectors. 

Finally, if concatenating the low dimensional data $\hat{\boldsymbol{x}} \in \mathbb{R}^{d}$ as columns of matrix $\hat{\boldsymbol{X}}$, we can solve $\hat{\boldsymbol{X}}$ by
\begin{equation}
\hat{\boldsymbol{X}} = \boldsymbol{P}^{\top}\boldsymbol{X}
\end{equation}

\subsection{Updating Low Dimensional Distance}
\label{ssec:update}

The low dimensional distance in $\mathbb{R}^d$ at the end of each iteration is updated as
\begin{equation}
M_{ij}^{\left(t+1\right)} \ = \ M_{ij}^{\left(t\right)} + \eta \left( \boldsymbol{z}_{ij}^{\top}\boldsymbol{P}^{\left(t+1\right)}\boldsymbol{P}^{\left(t+1\right)\top}\boldsymbol{z}_{ij} - M_{ij}^{\left(t\right)}\right)
\end{equation}
where $\eta \in (0,1]$ is the learning rate usually smaller than 0.1. 

Because in each iteration we have $J^{(t+1)}(\boldsymbol{P})>J^{(t)}(\boldsymbol{P})$, the iteration process should stop when the increment becomes tiny, and the value of $J{(\boldsymbol{P})}$ comes to stabilize. Then the finally resulting $\boldsymbol{P}$ is the projection matrix we want for the supervised linear projection of dimensionality reduction. 

\subsection{Diffusion Map}
\label{ssec:diffmap}

After the dimensionality reduction by linear projection $\boldsymbol{P}$, the Diffusion Map \cite{coifman2006diffusion}, which is non-linear and unsupervised, in the next step can achieve better performance than the case without our proposed projection. Results are shown in Section 5. 

Given a dataset with n points $\left\{\hat{\boldsymbol{x}}_1,\hat{\boldsymbol{x}}_2,\hat{\boldsymbol{x}}_3,...,\hat{\boldsymbol{x}}_n\right\}$, jumping to nearby points is more likely than to further points in the Markov random walk. Therefore, with $\sigma$ related to the variance of dataset $\hat{\boldsymbol{X}}$, the connectivity matrix $\boldsymbol{W}$ can be defined as the jumping probability between each two points under the Gaussian Kernel model as in (12), which is symmetric and positive semi-definite.

After the normalized transition probability matrix $\boldsymbol{T}$ is obtained as in (12), it is then applied to the eigen-decomposition. We get eigenvalues $\lambda$ in descending order, with right eigenvectors $\boldsymbol{\varphi}$. 
\begin{equation}
W_{ij} = \exp \left( -\frac{\left\|\hat{\boldsymbol{x}}_{i}-\hat{\boldsymbol{x}}_{j}\right\|_{2}^{2}}{\sigma^2} \right), \quad T_{ij} = \frac{W_{ij}}{\sum_{j} W_{ij}}
\end{equation}
Note that as $\left\|\hat{\boldsymbol{x}}_{i}-\hat{\boldsymbol{x}}_{j}\right\|_{2}^{2}$ decreases, $W_{ij}$ will increase, and so does the probability $T_{ij}$. In other words, when intra-class similarities are increased, so are the Markov random walk probabilities in the Diffusion Map, and vice versa.

At last, the data on the manifold discovered by the Diffusion Map after $t$ time steps has the form
\begin{equation}
\tilde{\boldsymbol{X}} = \left[\ \lambda_1^{t}\boldsymbol{\varphi}_1,\  \lambda_2^{t}\boldsymbol{\varphi}_2,\ ...,\ \lambda_d^{t}\boldsymbol{\varphi}_d\ \right]
\end{equation}
where each row of $\tilde{\boldsymbol{X}}$ is one data sample $\tilde{\boldsymbol{x}}$, and each $\lambda_l^{t} \boldsymbol{\varphi}_l$ with $l \in \left[1,d\right]$ is a column vector containing the values of all data samples at the coordinate $l$ in the lower dimensional space in $R^{\tilde{d}}$

In summary, the mapping $\hat{\boldsymbol{x}} \in \mathbb{R}^d \rightarrow \tilde{\boldsymbol{x}} \in \mathbb{R}^{\tilde{d}}$ with $\tilde{d} < d$ embeds the data to a lower dimensional Euclidean space, by the top $\tilde{d}$ eigenvalues and corresponding eigenvectors. At this stage, a classical classifier, such as linear SVM or KNN, is ready to be applied onto the found manifold of dimension $\tilde{d}$ for the classification task.

\section{DISCUSSION}
\label{sec:discuss}

Empirically, the objective function $J(\boldsymbol{P})$ in (4) usually reaches its maximum when all positive eigenvalues of $\boldsymbol{XLX}^{\top}$ are chosen, and the dimensionality $d$ is one less than the number of different classes.

Therefore, we update (4) using all positive eigenvalues as
\begin{equation}
J^{(t+1)}(\boldsymbol{P}) = J^{(t)}(\boldsymbol{P}) + \sum_{d}{\lambda_d} + \left(\left(1-\rho\right) \sum_{k=1}^K \lambda_k n_k - \rho n_o \right)
\end{equation}
where the last term $\left(\left(1-\rho\right) \sum_{k=1}^K \lambda_k n_k - \rho n_o \right)$ in (14) is constant for all iterations, with no effect on convergence.

It can also be seen that, as long as the eigenvalues selected are all positive, the objective function will always increase. In other words, the relation $J^{(t+1)} \left(\boldsymbol{P}\right) > J^{(t)}\left(\boldsymbol{P}\right)$ will be guaranteed.

\section{EXPERIMENT RESULTS}
\label{sec:result}

Numerical experiments have been performed for image and video classifications with two popular labeled benchmark dataset. First, the classical image datasets have been tested for our proposed linear projection $\boldsymbol{P}$ in (9) to demonstrate the optimization of intra-class and inter-class similarities. Then, the human action sequencing dataset, which contains an inherent low dimensional non-linear manifold, has been tested on the method we proposed, as the effective way of improving classification accuracy. 

\subsection{Optimization of intra-class and inter-class similarities}
\label{ssec:exp1}

For image dataset Caltech-101 \cite{fei2007learning} and 15 Scenes \cite{lazebnik2006beyond}, where underlying manifolds are not obvious, we will demonstrate graphically that our proposed linear projection $\boldsymbol{P}$ in equation (9) can indeed increase intra-class similarities and decrease inter-class similarities. 

High dimensional features are extracted by linear spatial pyramid matching (SPM) \cite{yang2009linear} from raw grayscale images. Specifically, the SIFT \cite{lowe1999object,vedaldi2010vlfeat} descriptors are extracted from 40$\times$40 pixel patches with step size 8 pixels. Then, the dictionary for calculating the sparse coding features is trained through a regularized sparse coding technique \cite{lee2007efficient} with k-nearest neighbors to perform approximate max pooling, and the feature vector has size 21504$\times$1 for each image. PCA \cite{jolliffe2002principal,abdi2010principal} is then applied from 21504 feature dimensions to maximum possible feature dimensions, in order to compress the original data for more efficient calculations. Data now is the input $\boldsymbol{x} \in \mathbb{R}^D$.

The Caltech-101 contains 102 classes and 9144 images. For each class, 30 training samples and at most 40 testing samples are used. The 15-Scenes contains 15 classes and 4485 images, with 100 training samples and at most 150 testing samples used in each class.

Then, our dimensionality reduction method is ready to be applied to get the low dimensional data $\hat{\boldsymbol{x}} \in \mathbb{R}^d$.

\begin{figure}[htb]
\begin{minipage}[b]{1.0\linewidth}
  \centering
  \centerline{\includegraphics[width=9cm]{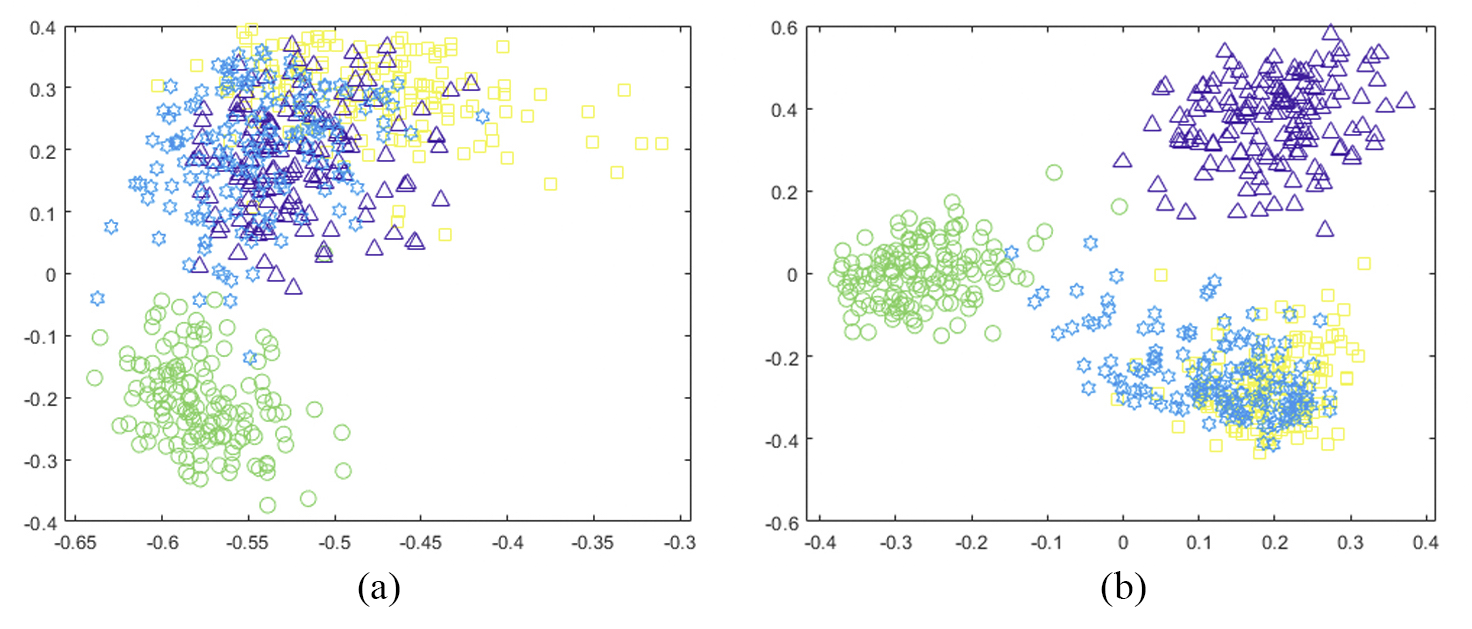}}
\end{minipage}
\vspace{-0.7cm}
\caption{(a) Data in high dimension (b) data after linear projection $\boldsymbol{P}$}
\label{fig:figure1}
\end{figure}

Figure 1 is a typical example from 15-Scenes, showing the first two dimensions and four classes. It can be seen that samples within the same class get more clustered, while clusters of different classes get more separated. Therefore, the pairwise local kernel distances are optimized based on the label information, and so are the similarities defined in equation (1), as expected.

Linear projection $\boldsymbol{P}$ is able to improve classification accuracy compared with high dimensional data, and also outperforms PCA. In this case, Linear SVM \cite{cortes1995support} will be used as the classifier, as suggested in \cite{yang2009linear}, and the experiment is run 10 times with different random selected training and testing samples. Results are shown in Table 1.

\begin{table}[h]
\centering
\begin{tabular}{ | c | c | c | c | }
\hline
\thead{Method}  & \thead{High Dimension} & \thead{PCA} & \thead{$\boldsymbol{P}$}\\
\hline
\makecell{Caltech-101} & \makecell{73.673$\pm$0.909} & \makecell{73.713$\pm$0.832} & \makecell{74.726$\pm$0.982}\\
\hline
\makecell{15 Scenes} & \makecell{76.131$\pm$1.091} & \makecell{75.826$\pm$0.999} & \makecell{81.536$\pm$0.644}\\
\hline
\end{tabular}
\caption{Average accuracy and standard deviations (\%)}
\vspace{-0.4cm}
\end{table}

\begin{figure*}[t]
\begin{minipage}[b]{1.0\linewidth}
  \centering
  \centerline{\includegraphics[width=18cm]{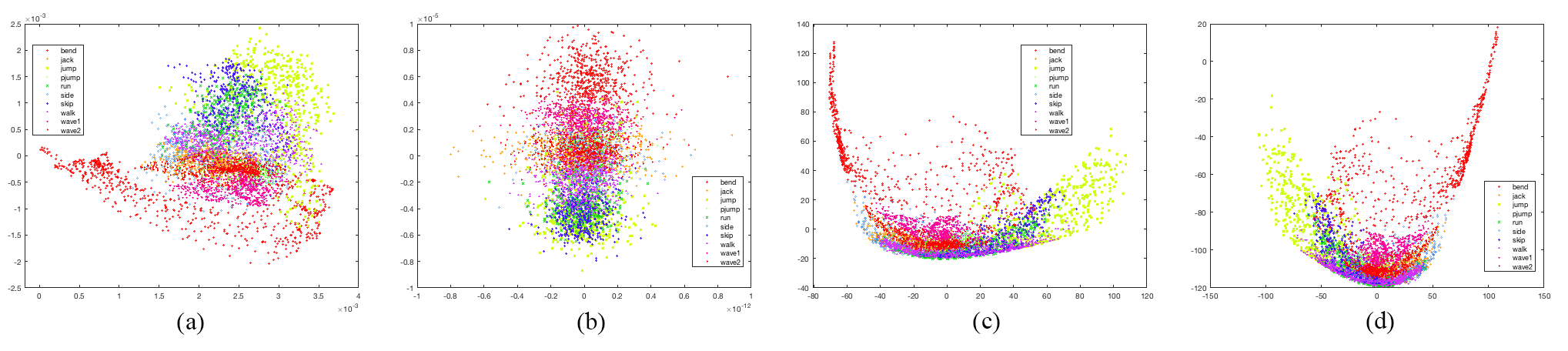}}
\end{minipage}
\vspace{-0.8cm}
\caption{Low dimensional 2D manifold: (a) PCA (b) LDA (c) DM (d) linear projection $\boldsymbol{P}$ + DM}
\label{fig:figure2}
\vspace{-0.2cm}
\end{figure*}

\begin{figure*}[t]
\begin{minipage}[b]{1.0\linewidth}
  \centering
  \centerline{\includegraphics[width=18cm]{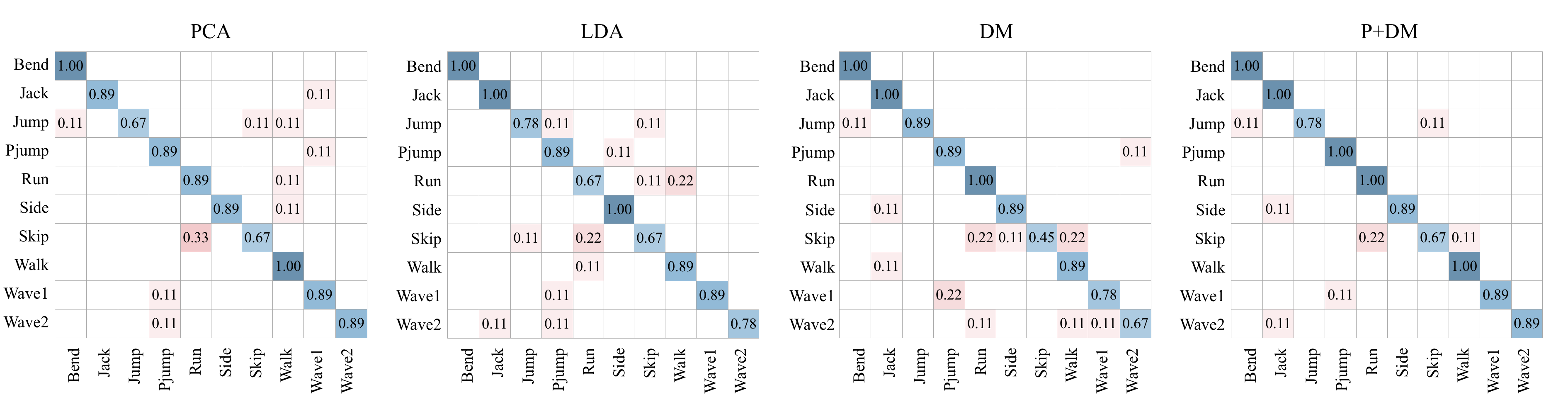}}
\end{minipage}
\vspace{-0.6cm}
\caption{Confusion table for different dimensionality reduction methods.} \vspace{-0.2cm}
\label{fig:figure3}
\end{figure*}

\subsection{Improvement of Diffusion Map's classification accuracy}
\label{ssec:exp2}

We implemented the experiment again on the human action dataset \cite{gorelick2007actions} to test our algorithm. There are 9 human actions from 10 persons, each of which is a video sequence consists of a different number of image frames. Actions include bending (bend), jumping jack (jack), jumping-forward-on-two-legs (jump), jumping-in-place (pjump), running (run), galloping sideway (side), skipping (skip), walking (walk), waving-one-hand (wave1), and waving-two-hands (wave2).

The silhouette of each image frame is used in our experiment, due to its efficiency to represent human actions \cite{gorelick2007actions}. Then the 2D silhouette image $\boldsymbol{I}$ is transformed into a 1D feature vector $\boldsymbol{x}$ by $\mathcal{R}-$transforms \cite{tabbone2006new}. This $\boldsymbol{x}$ is the input data to our algorithm. Define $I\left(i,j\right)$ as the pixel in $\boldsymbol{I}$ at position $\left(i,j\right)$, and $\rho$, $\theta$ as displacements and angles in the Radon Transform \cite{helgason1999radon}, we have
\begin{equation}
\begin{aligned}
&T\left(\boldsymbol{I}, \rho, \theta\right)=\sum_{i} \sum_{j} I(i, j) \delta(i \cos \theta+j \sin \theta-\rho)
\\
&\boldsymbol{x} = \mathcal{R}\left(\boldsymbol{I}, \theta\right)=\sum_{\rho} T^{2}\left(\boldsymbol{I}, \rho, \theta\right) / \sum_{\rho} \sum_{\theta} T^{2}(\rho, \theta)
\end{aligned}
\vspace{-0.1cm}
\end{equation}

So far, the high dimensional feature $\boldsymbol{x} \in \mathbb{R}^D$ is ready to be applied with linear projection $\boldsymbol{P}$ to get $\boldsymbol{\hat{x}} \in \mathbb{R}^d$, followed by the Diffusion Map to get data $\boldsymbol{\tilde{x}} \in \mathbb{R}^{\tilde{d}}$ in a lower dimensional manifold. 

We have compared our methods with other classical dimensionality reduction algorithms, including Linear Discriminate Analysis (LDA) \cite{fisher1936use,mclachlan2004discriminant}, Principle Component Analysis (PCA) \cite{jolliffe2002principal,abdi2010principal}, and DM without $\boldsymbol{P}$ in advance (keeping the parameters for DM the same as in DM with $\boldsymbol{P}$). The resulting data distributions in the final low dimensional manifolds are shown in Figure 2. In a comparison between Figure 2 (c) and (d), it can be seen that data within the same class gets more clustered in (d), where the supervised linear projection $\boldsymbol{P}$ is applied before DM.

In classification by KNN, each frame in the testing video is assigned to the most frequent label of its $k$ nearest neighbours in the training data, and the label of the whole video is determined by the majority vote of all its frames. Nine-fold cross-validation is adopted in the evaluation. Each time all sequences of a single person are the testing data, while the remaining are the training data. 

The comparison of classification accuracy is shown in Table 2, as well as the confusion table in Figure 3. The experiment shows that our proposed linear projection $\boldsymbol{P}$, used before the Diffusion Map, helps promote the overall classification accuracy.
\begin{table}[htb]
\centering
\begin{tabular}{ | c | c | c | c | c | }
\hline
\thead{Method} & \thead{PCA} & \thead{LDA} & \thead{DM} & \thead{$\boldsymbol{P}$+DM}\\
\hline
\thead{Accuracy} & \makecell{86.7} & \makecell{85.6} & \makecell{87.8} & \makecell{91.1}\\
\hline
\end{tabular}
\caption{Overall classification accuracy with different methods (\%)}
\vspace{-0.5cm}
\end{table}

\section{CONCLUSION}
\label{sec:conclu}

In this paper, we proposed a supervised dimensionality reduction approach to optimize the high dimensional data distributions based on label information, so that the unsupervised Diffusion Map, as the further dimensionality reduction in the next step, can find better low dimensional manifolds, where higher classification accuracy can be achieved. The proposed approach searches for linear projections so that the intra-class similarities are maximized and the inter-class similarities are minimized. Meanwhile, it removes redundant information contained in the original dataset to promote computation efficiencies. Image and video classification experiments on the benchmark dataset \cite{gorelick2007actions} using the extracted low dimensional features show that the classification accuracy with Diffusion Map could be improved more than before. In our future work, we would like to investigate how the neighborhood size in (3) could affect the extracted features, and test the algorithm on other large-scale datasets.

\section{ACKNOWLEDGEMENT}
\label{sec:acknow}

The authors would like to thank Prof. Yoram Bresler and Dr. Shuai Huang at the Coordinate Science Laboratory, UIUC for their guidance and support in helping us complete the work.


\bibliographystyle{IEEEbib}
\bibliography{refs}

\begin{thebibliography}{10}

\bibitem{Bellman:1957}
Richard Bellman,
\newblock {\em Dynamic Programming},
\newblock Princeton University Press, Princeton, NJ, USA, 1 edition, 1957.

\bibitem{trunk1979problem}
Gerard~V Trunk,
\newblock ``A problem of dimensionality: A simple example,''
\newblock {\em IEEE Transactions on Pattern Analysis \& Machine Intelligence},
  , no. 3, pp. 306--307, 1979.

\bibitem{jolliffe2002principal}
Ian~T Jolliffe,
\newblock ``Principal components in regression analysis,''
\newblock {\em Principal component analysis}, pp. 167--198, 2002.

\bibitem{abdi2010principal}
Herv{\'e} Abdi and Lynne~J Williams,
\newblock ``Principal component analysis,''
\newblock {\em Wiley interdisciplinary reviews: computational statistics}, vol.
  2, no. 4, pp. 433--459, 2010.

\bibitem{fisher1936use}
Ronald~A Fisher,
\newblock ``The use of multiple measurements in taxonomic problems,''
\newblock {\em Annals of eugenics}, vol. 7, no. 2, pp. 179--188, 1936.

\bibitem{mclachlan2004discriminant}
Geoffrey McLachlan,
\newblock {\em Discriminant analysis and statistical pattern recognition}, vol.
  544,
\newblock John Wiley \& Sons, 2004.

\bibitem{gorelick2007actions}
Lena Gorelick, Moshe Blank, Eli Shechtman, Michal Irani, and Ronen Basri,
\newblock ``Actions as space-time shapes,''
\newblock {\em IEEE transactions on pattern analysis and machine intelligence},
  vol. 29, no. 12, pp. 2247--2253, 2007.

\bibitem{coifman2006diffusion}
Ronald~R Coifman and St{\'e}phane Lafon,
\newblock ``Diffusion maps,''
\newblock {\em Applied and computational harmonic analysis}, vol. 21, no. 1,
  pp. 5--30, 2006.

\bibitem{cortes1995support}
Corinna Cortes and Vladimir Vapnik,
\newblock ``Support-vector networks,''
\newblock {\em Machine learning}, vol. 20, no. 3, pp. 273--297, 1995.

\bibitem{lowe1999object}
David~G Lowe et~al.,
\newblock ``Object recognition from local scale-invariant features.,''
\newblock in {\em iccv}, 1999, vol.~99, pp. 1150--1157.

\bibitem{vedaldi2010vlfeat}
Andrea Vedaldi and Brian Fulkerson,
\newblock ``Vlfeat: An open and portable library of computer vision
  algorithms,''
\newblock in {\em Proceedings of the 18th ACM international conference on
  Multimedia}. ACM, 2010, pp. 1469--1472.

\bibitem{fei2007learning}
Li~Fei-Fei, Rob Fergus, and Pietro Perona,
\newblock ``Learning generative visual models from few training examples: An
  incremental bayesian approach tested on 101 object categories,''
\newblock {\em Computer vision and Image understanding}, vol. 106, no. 1, pp.
  59--70, 2007.

\bibitem{lazebnik2006beyond}
Svetlana Lazebnik, Cordelia Schmid, and Jean Ponce,
\newblock ``Beyond bags of features: Spatial pyramid matching for recognizing
  natural scene categories,''
\newblock in {\em 2006 IEEE Computer Society Conference on Computer Vision and
  Pattern Recognition (CVPR'06)}. IEEE, 2006, vol.~2, pp. 2169--2178.

\bibitem{yang2009linear}
Jianchao Yang, Kai Yu, Yihong Gong, Thomas~S Huang, et~al.,
\newblock ``Linear spatial pyramid matching using sparse coding for image
  classification.,''
\newblock in {\em CVPR}, 2009, vol.~1, p.~6.

\bibitem{lee2007efficient}
Honglak Lee, Alexis Battle, Rajat Raina, and Andrew~Y Ng,
\newblock ``Efficient sparse coding algorithms,''
\newblock in {\em Advances in neural information processing systems}, 2007, pp.
  801--808.

\bibitem{tabbone2006new}
Salvatore Tabbone, Laurent Wendling, and J-P Salmon,
\newblock ``A new shape descriptor defined on the radon transform,''
\newblock {\em Computer Vision and Image Understanding}, vol. 102, no. 1, pp.
  42--51, 2006.

\bibitem{helgason1999radon}
Sigurdur Helgason and S~Helgason,
\newblock {\em The radon transform}, vol.~2,
\newblock Springer, 1999.

\end{thebibliography}

\end{document}